\documentclass[aps,onecolumn,10pt]{revtex4-2}
\usepackage{graphicx}  
\usepackage{dcolumn}   
\usepackage{bm}        
\usepackage{amssymb}
\usepackage{amsmath} 

\usepackage{xcolor}
\usepackage{hyperref}
\usepackage{ulem} 

\newcommand{\Force}{\mathcal{A}}

\newcommand{\dd}{\textrm{d}}
\newcommand{\tomega}{\tilde{\omega}}

\begin{document}

\title{Large-Scale Turbulent Pressure Fluctuations Revealed by Ned Kahn's Artwork}

\author{J.~Zhang}
\author{S.~Perrard}
\affiliation{PMMH, CNRS, ESPCI Paris, Universit\'e PSL, Sorbonne Universit\'e, Universit\'e de Paris, F-75005, Paris, France}
\date{\today}

\begin{abstract}

We investigate the dynamics of pendulum chains immersed in turbulent boundary layers. We combine qualitative video analysis of the artist Ned Kahn's \textit{kinetic façades} of buildings, and laboratory experiments on a unidimensional weakly coupled chain of pendulums. We performed analysis of both the façades and the laboratory scale model. We show that pendulum waves travelled in the direction of the wind. These waves originate from the excitation by the spatio-temporal pressure fluctuations. We identified two dispersion relation branches. The first branch corresponds to a resonant response of each pendulum at its natural {frequency of oscillations}. The second branch corresponds to an excitation by the advected pressure fluctuations along the chain. Using local pressure sensors, we show a quantitative agreement between the convection velocity of the pendulum chain waves and the pressure fluctuations. Eventually, we propose a model in Fourier space to describe the magnitude of each branch. We show that at a small wind speed, the pendulum response is dominated by the resonance at their natural frequency. At larger wind speed, the response becomes dominated by the advected pressure fluctuations.

\end{abstract}

\maketitle

\section{Introduction} \label{sec:intro} 
\begin{figure}[htbp]
  \includegraphics[width=1\columnwidth]{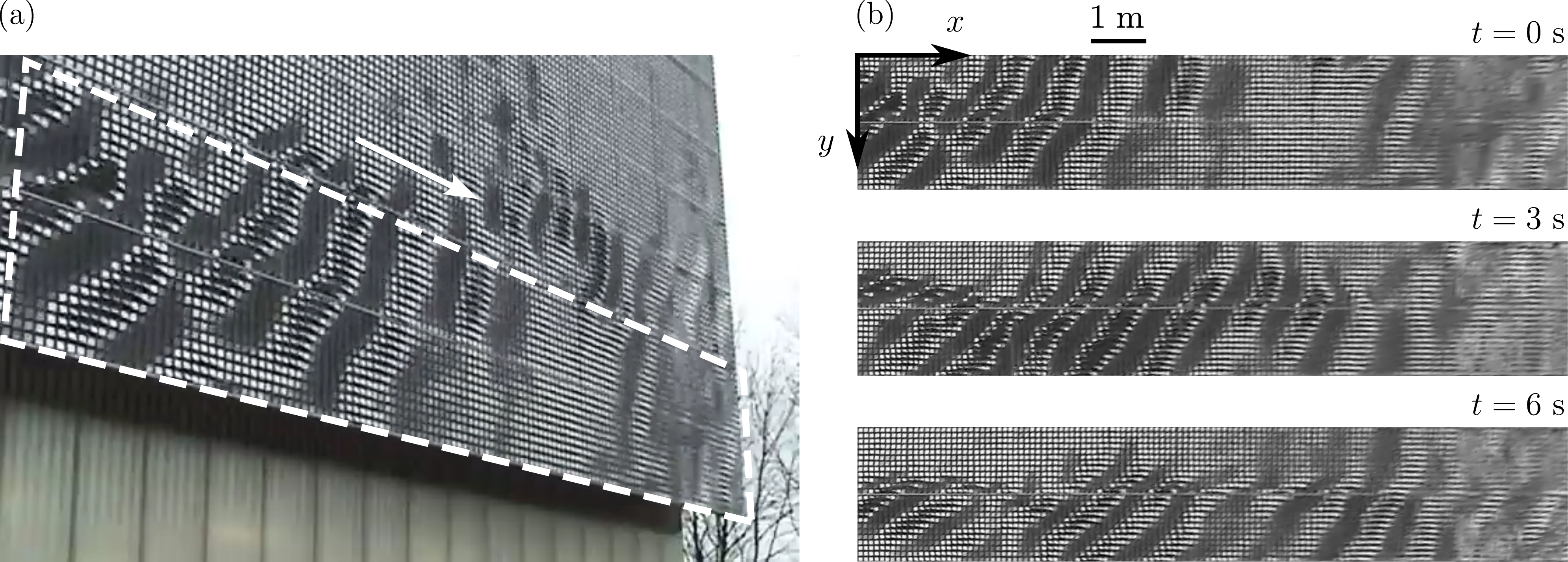}
  \caption{\label{fig:terrain} a: snapshot of wave patterns on Ned Kahn's kinetic façade with the region of interest delimited by a dashed-border quadrilateral; b: distortion-corrected image of the region of interest in three sequence frames.}
\end{figure}

Ned Kahn is an American artist who constructs numerous exhibits inspired by the ephemera of nature. Amongst his works is the kinetic façade, a regular assembly of small aluminum plates hinged to wall-attached grid covering the entire facade of buildings in various countries (US, Scotland, Netherland, Switzerland, France). As the wind blows along the wall, the plates oscillate freely, creating propagative wave-like large-scale patterns (fig.~\ref{fig:terrain}). The generation of such complex structures could be the result of numerous mechanisms, such as fluid-structure instabilities observed in flapping wings~\cite{shelley2011flapping}, flags~\cite{ristroph2008anomalous,connell2007flapping} and canopies~\cite{de2008effects}, or bistability of pendulums in turbulent flows~\cite{Gayout_2021}. Recent study on an air flow above a viscous liquid surface otherwise quiescent~\cite{paquier2016viscosity} suggests that these patterns might be the signature of the wall-pressure fluctuations induced by the turbulent boundary layer near the plate surface (TBL)~\cite{perrard2019turbulent}.

Turbulent pressure fluctuations are ubiquitous in natural and industrial fluid systems. In the scenario of a TBL adjacent to a flexible solid interface, the spatiotemporal pressure fluctuations interact with the eigenmodes of the structure and contribute to the resonance-induced vibration in many industrial applications leading to a crack propagation in aircraft wings or pressure vessel nozzles~\cite{blake2017mechanics}. In atmospheric TBL, knowledge of the spatiotemporal pressure fluctuations at large scale~\cite{He_2017} are also key elements in wind farm design to minimize the global power fluctuations of a plant affected by long-range fluctuations. These fluctuations have been characterized by two-point measurements in the far wake region~\cite{sorensen2002wind, wetz2023analyses, vermeer2003wind}. However, simultaneous measurements of both the spatial and temporal components of the turbulent structures remain still a long-standing challenge. Numercially, it has been shown~\cite{del2009estimation} that the wall pressure fluctuations travel at a speed $U_c(k)$ smaller than the bulk mean speed, which depends on the mode wavenumber $k$.

In this paper, we aim to characterize the waves that are observed on Ned Kahn's kinetic façades. To do so, we performed the analysis of kinetic facades video recording. To test the sensitivity in wind speed, we design a reduced one-dimensional model, made of a chain of weakly coupled pendulums. We study the response of the pendulum chain to turbulent fluctuations as a function of the wind speed. We perform in particular a component-by-component analysis in the space and time Fourier space of the pendulum chain.


\section{Survey data collection} \label{sec:Kineticfacade} 

To characterize the dynamic patterns of the kinetic façades, we rely on video extracts recorded by amateurs. A total of 18 videos taken on six different facades have been gathered from YouTube.com and Vimeo.com, with typical frame rates ranging from $25$ to $29$ Hz. Most facades are covered with square plates of flapping length $l$ ranging from $51$~mm to $127$~mm, and separated by a distance ranging from $1.2$ to $1.6$ times the flapping length. The dimensions of the plates were collected by direct communications with Ned Kahn and technical services of the corresponding buildings. The spatial dimensions were also used to recover the scale factor from the amateur videos. To account for the variability in viewing angles of the camera, we apply a quadratic transformation mapping the quadrilateral distorted plate images to rectangles.  A video snapshot and a corrected image of the facade from the \textit{Swiss Science Center Technorama}~\citep{technoramaF} are shown in fig.~\ref{fig:terrain} with wavy patterns at three different instants. These moving patterns are observed with the reflecting light on the oscillating plates that leads to a change of pixel brightness as the pendulum angle changes. The $x$ axis is parallel to the ground surface, and the $y$ axis is vertical, pointing downward. To facilitate the data analysis, image sequences are horizontally flipped to ensure that the patterns propagate always toward the positive $x$ direction. 

\begin{figure}[htbp]
  \includegraphics[width=1\columnwidth]{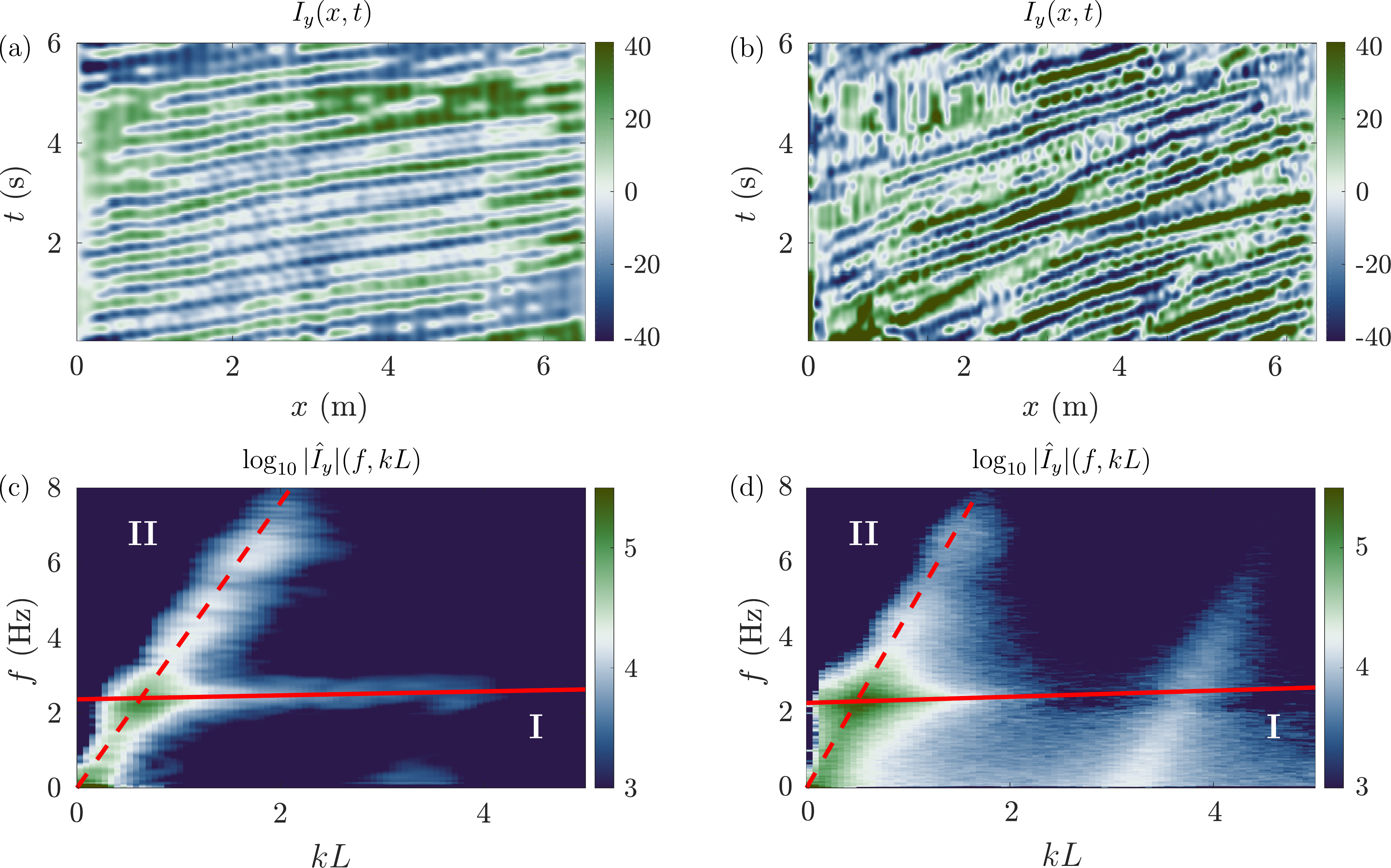}
  \caption{\label{fig:spec_terrain} Pattern dynamics for two example facades~(left: \textit{Glacial facade}~\citep{glacialF}, right: \textit{Digitized Field}~\citep{digitizedF}). (a,b) spatiotemporal charts of the image pixel intensity for an image row along $x$. (c,d) spatiotemporal spectrum of the image pixel intensity along $x$, averaged over the $y$ direction {with color-bars are shown with logarithmic scale.}}
\end{figure}

In the absence of quantitative measure of the pendulum's instantaneous angle of rotation $\theta(x,y,t)$, we rely on the temporal variation of pixel brightness $I(x,y,t)$ by assuming a monotonous relation between $\theta$ and $I$. By doing so, we can qualitatively study the spatio-temporal behaviors of the moving patterns. As the videos are taken in natural conditions, both the wind amplitude and its direction changes over time (fig.~\ref{fig:terrain}b). The wavy patterns generated propagate mainly in $x$ direction but also exhibit some propagative bursts with a vertical component. From our analysis, we speculate that this fluctuating direction of propagation comes from the fluctuations of the wind direction, but we have not studied in details this phenomenon. To study the propagation along $x$, we truncated the video clip to focus on video extract in which the waves propagate mainly along $x$. We decompose the image sequence into two-dimensional space vectors $I_y(x,t)$ containing the pattern propagation in $x$ and $t$ for each row in $y$. To examine the frequency–wave number spectrum of the kinetic façade patterns, we perform the two-dimensional discrete Fourier transform in space and time that converts $I_y(x,t)$ from physical space into spectral space $\hat{I}_y(k,\omega)$:

\begin{gather}
    \hat{I}_y(k,\omega) = \mathcal{F} \left \{ I_y(x,t) \right \} = \int d^2x dt I_y(x,t) e^{-i(kx-\omega t)}\label{eq:Fourier1} \\
     I_y(x,t) = \mathcal{F}^{-1}\left \{ \hat{I}_y(k,\omega) \right \}=(2\pi)^{-3} \int d^2 k d\omega \hat{I}_y(k,\omega) e^{i(kx-\omega t)}
    \label{eq:Fourier2}
\end{gather}

with $k$ the wave number of the patterns in the $x$ direction and $\omega$ the angular frequency of the plate oscillation. Figure~\ref{fig:spec_terrain} shows the pattern propagations in physical space ($x, t$) and the corresponding frequency-wave number spectra averaged over the $y$ direction for two facades~(left: \textit{Glacial facade}~\citep{glacialF}, right: \textit{Digitized Field}~\citep{digitizedF}). The wave number $k$ is made dimensionless using the spacing $L$ between two adjacent pendulums.

In Fourier space, we found that the pixel brightness is localized along two main branches. The branch II is located around $f = U_c k/(2 \pi)$ (dashed red line), where $U_c$ is a typical velocity. This velocity corresponds to the convection of pendulum fluctuations at a constant speed $U_c$, almost independent of the wave number. 

The branch I is nearly horizontal, located around a typical frequency $f_0$, that corresponds to the natural oscillation frequency of each pendulum given by $f_0=\sqrt{1.5g/l}/(2\pi)$, where $l$ is the flapping length. This first branch can be interpreted as a resonant response at the natural flapping frequency of each pendulum to the turbulent fluctuations. 
In the absence of coupling between adjacent pendulum, we expect this first branch to be horizontal, corresponding to $f=f_0$ for all wave numbers $kL$. 
However, we found a slight increase of $f$ with $kL$ in all cases. This weak trend can be interpreted as a pendulum coupling induced by the wind blow. Similar behaviors have been observed for flags in an air flow. In this context, the inertia of the air flow brings about a linear contribution $m_AU\partial^2A/\partial x^2$ to the flag normal force balance, known as an added stiffness effect~\cite{de2001fluides, ramananarivo2014propulsion}. The mass of the otherwise air volume occupied by the flag is $m_A$ and $A$ is local out-of-plane amplitude of the flag. 
This term being proportional to the local curvature of the flag interface and the air flow velocity, tends to destabilize the flag motion. However, here we observe an increase of the natural frequency with the wave number, reminiscent of a stabilizing effect. Overall, this wind-induced coupling effect remains small, with less than 10 $\%$ of frequency increase at $kL=1$. It can be attributed to the presence of holes between pendulums, that prevent the build-up of large pressure differences between the two faces. We can therefore consider that the wave propagation is marginally affected. An image of branch II appears at larger wave numbers ($kL>2$) {in fig.~\ref{fig:spec_terrain}d}, and it originates from the secondary maximum at higher wave number of the Fourier transform of the rectangular plates brightness. For all the analysed videos, we have observed these two main branches.

To summarize, spatiotemporal spectral analysis on kinetic façades suggest that the facades are excited by two mechanisms: a resonant response at all wave numbers around $f=f_0$ (branch I) and a direct response to turbulent fluctuations traveling at constant convection speed $U_c$ (branch II). The maximum response is reached at the intersection between the two branches, which meets both criteria: the pendulum responds to spatiotemporal turbulent fluctuations that excite their natural frequency, corresponding to a wave number $k_{max} = 2\pi f_0/U_c$. To investigate the origin of the plate motions and better identify the mechanism at play, we built a one dimensional laboratory model.


\section{Laboratory set-up} \label{sec:setup} 
\begin{figure}
  \includegraphics[width=0.8\columnwidth]{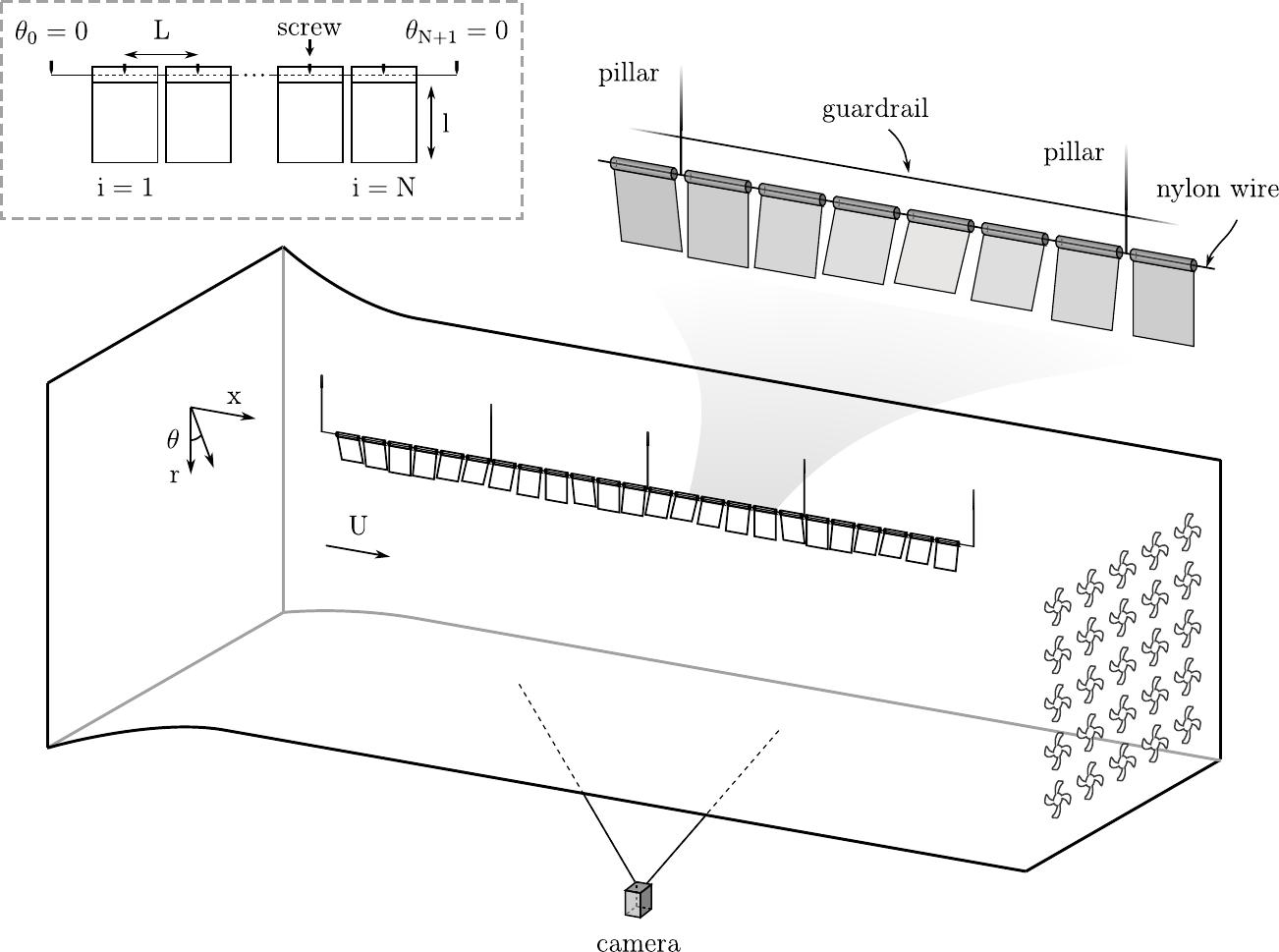}
  \caption{\label{fig:schema} Sketch of the experimental set-up, composed of a thin plate chain elastically coupled, and placed in the measuring section of the wind tunnel.}
\end{figure}

We design a one-dimensional experimental model at the laboratory scale, composed of a chain of pendulum plates, as sketched in fig.~\ref{fig:schema}. The chain measures $1.1$~m long, and it consists of $N=36$ 3D-printed pendulum thin plates with flapping length $l=48.5$~mm; width $w=28.5$~mm, thickness $h=1$~mm, mass $m=2.33$~g and density $\rho_p=1040$~kg/$m^3$, equally spaced by a distance $L=32.5$~mm. We use screws to attach each pendulum center to a nylon wire, sufficiently thin to prevent significant mechanical coupling between plates. To limit the chain warping induced by gravity, a sustaining pillar holds the wire every six pendulums while leaving the wire free to rotate. At both chain ends the wire is clamped onto the pillar's tip, imposing a zero rotating angle boundary condition~(fig.~\ref{fig:schema} inset) at a distance $L$ from the first and last plates. A guardrail wire spanning the whole length of the chain is attached to the pillars, this prevents the pendulums to make a full lap around the rotation axis thus keeps the rotation angle of each pendulum limited between $-\pi$ and $\pi$. The pendulum connection using the nylon wire of $a = 0.2$ mm in diameter introduces a weak elastic coupling constant between adjacent plates. 
In the approximation of zero elastic coupling constant, each pendulum is free to oscillate at a natural frequency $\omega_0=\sqrt{1.5g/l}=17.4~\mathrm{rad./s}$. In the absence of any external applied force and with negligible mechanical coupling between plates the motion of each pendulum follows:
\begin{equation}
  \partial_{tt}\theta_i + \lambda \partial_t \theta_i + \omega_0^2\sin\theta_i= 0.
  \label{eq:force_balance}
\end{equation}
In practice, we limit ourself to pendulum angles of the order of 0.4 rad. The equation of motion can therefore be linearized, $\sin \theta_i \sim \theta_i$. As a consequence, we will assume in particular that the {frequency of oscillations} $\omega_0$ is independent of the pendulum angle. 
In the absence of flow, the damping coefficient $\lambda$ is of the order of $0.3$ s$^{-1}$, which is much smaller than the natural angular frequency $\omega_0$. However, this damping should never be neglected, as it limits the amplitude growth near the resonance.




The pendulum chain is immersed in a turbulent flow, generated by an open circuit suction wind tunnel. The chain is placed at the symmetry plane of the wind tunnel. We limit the free-stream wind speed $U$ between 0 to $4.6$~m/s. Above 4.6~m/s, the pendulum starts to swing around the wire episodically. The wind tunnel has a measuring section of 720 $\times$ 720 $\times$ 1400 mm and a contraction ratio of 1.4. This low contraction design leads to a significant turbulent intensity in the incoming flow~\cite{bell1988contraction} and the grid-free open entrance makes the inlet flow susceptible to the room ambient turbulence where the large-scale structures are preferably prominent. All experiments were performed in a closed room, with no external current. Independent measurements on the turbulent statistics and the convection velocity of the pressure fluctuations in the wind tunnel were performed using a hot-wire probe and two pressure sensors. The hot-wire probe was calibrated using the King's law and a resistor anemometer Testo 425.

We define the typical {velocity fluctuations} as $u'= \sqrt{\langle u_x - \langle u_x \rangle \rangle}$. The turbulent intensity $u'/U = 0.094 \pm 0.01$ is constant over the full range of explored wind speed. We estimate the integral length scale $L_{int}$ from the temporal longitudinal auto-correlation function of velocity fluctuations, using Taylor hypothesis. We found that $L_{int}$ increases linearly at low wind speed, ranging from $L_{int} = 2.8$ cm for $U$=1.5~m/s to $L_{int} = 6.2$cm for $U$=4.6~m/s. This increase of the integral length scale with the wind speed is anomalous. It can be attributed to the natural injection of velocity fluctuations from upstream in the absence of a meshed grid. With a grid placed upstream, we would have expected an approximately constant integral length scale with wind speed. A first estimate of the dissipation rate $\epsilon$ can be obtained from the dissipation law $\epsilon = C u'^3/L_{int}$ for homogeneous and isotropic turbulence\cite{Pope_2000, vassilicos2015dissipation}, with $C\approx0.5$. We find a quadratic increase of $\epsilon$ with wind speed, with an upper value of about $\epsilon$ = 0.8 m$^2$/s$^{-3}$ for $U$= 4.6~m/s. From the second longitudinal structure function $S_2(r) = \langle (u(x+r,t)-u(x,t))^2 \rangle_t$, we estimate the range of inertial scale that follows K41 scaling\cite{kolmogorov1941local, Pope_2000}, $S_2(r) = C_2\epsilon^{2/3} r^{2/3}$, with $C_2 \approx 2$ from the literature\cite{Pope_2000}. We do observe a plateau for the compensated structure function $(S_2(r)/C_2)^{3/2} r^{-1}$ at all wind speeds. For $U=4.6$~m/s, we extract the value of the dissipation rate $\epsilon = 0.72$ m$^2$/s$^{-3}$ from the plateau. This value is compatible with our estimate from the large scale dissipation law. Assuming isotropic turbulence, we can also estimate the Taylor microscale $\lambda_T$, the scale at which the velocity gradients are maximum, from the dissipation rate and the velocity fluctuations $u'$. We have $\lambda_T = (15 \nu/\epsilon)^{1/2} u' \approx 8$~mm, and a corresponding Taylor Reynolds number $Re_\lambda = u' \lambda_T/\nu \approx 250$ for $U = 4.6$~m/s and $Re_\lambda = 140$ for $U = 1.5$ m/s. Overall, this wind tunnel facility provides a simple flow, homogeneous in the central region, to immerse objects of centimetric sizes, of size comparable with the integral length scale. Note that the pendulum chain length is much larger than the integral length scale: the pendulum chain oscillations will probe structures that are larger than the integral scale.

\begin{figure}
  \includegraphics[width=\columnwidth]{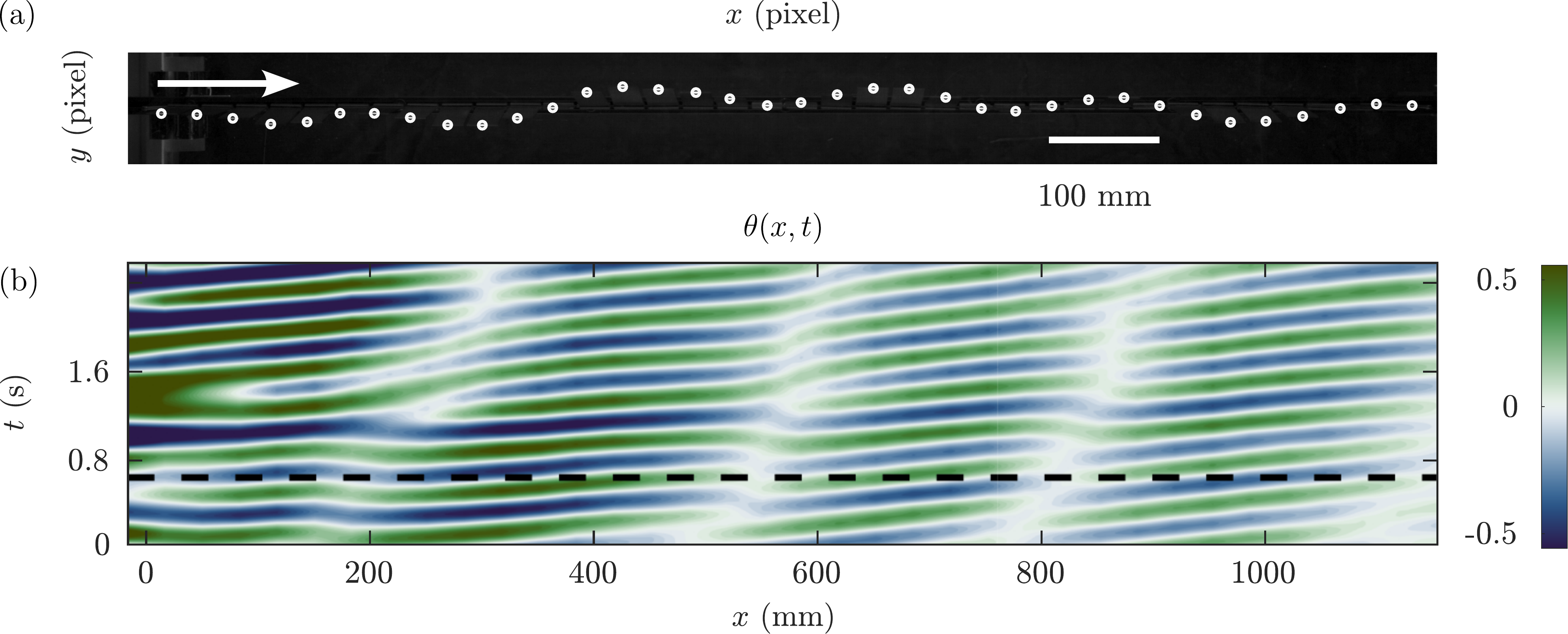}
  \caption{\label{fig:timeshot} a: snapshot of the oscillation of pendulums under a mean wind velocity $U=3.38$~m/s. The pendulum's tip positions are highlighted with white circles for visibility; b: spatiotemporal chart of the pendulum oscillation after image processing, with the instant illustrated above (black dashed line).}
\end{figure}

Immersed at the center of the channel, the pendulum motions are recorded from below and a white spot painted on each pendulum tip is used to track the instantaneous pendulum angles of orientation. For each wind speed, $90000$ images are recorded at a frame rate of $300.03$ images per second. From the sets of images, a spot-recognition code with sub-pixel accuracy is used to extract the pendulum angles $\theta_i(t)$. Figure~\ref{fig:timeshot} shows respectively a snapshot of the pendulum chain, and the spatiotemporal diagram of the inclination angles $\theta_i(t)$ for a wind speed of 3.38 m/s. Similarly to Ned Kahn's facade, we observe the propagation of patterns at an almost constant speed, traveling downstream along the pendulum chain. 

\begin{figure}
  \includegraphics[width=1\columnwidth]{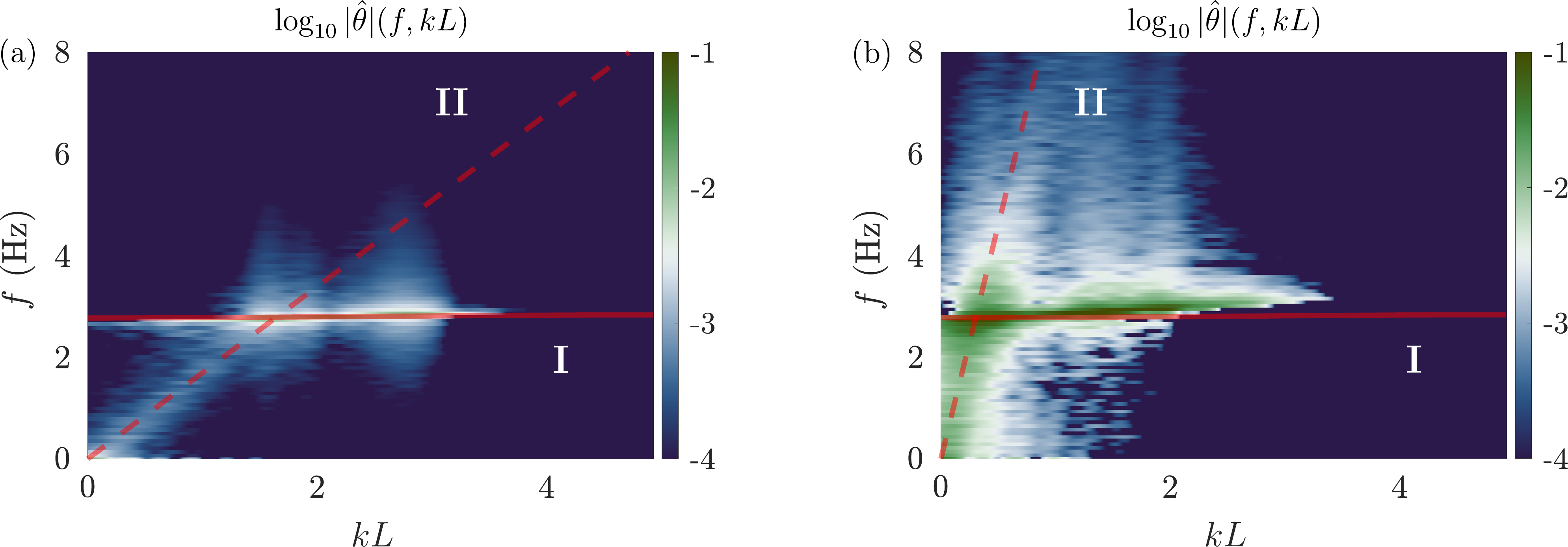}
  \caption{\label{fig:RD2D} Spatiotemporal spectrum of pendulum oscillations $|\hat{\theta}|(f, kL)$ {in logarithmic scale} (a: $U=1.16$~m/s; b: $U=3.16$~m/s), with the natural frequency constant (red dashed line) and the best fit of constant convection speed $\omega = U_c k$ (solid red line).}
\end{figure}

We measure each $\theta_i$ as a function of time, and we compute the Fourier transform $\hat{\theta}(k,f)$ of $\theta$ both in $x$ and time. The colormaps of $\hat{\theta}(k,f)$ are shown in fig.~\ref{fig:RD2D} for two different wind speeds of $U=1.16$~m/s and $U=3.16$~m/s. With the weakly coupled pendulum chain, we observe the same two main branches previously observed on the building facades.

The branch I corresponds to the propagation of waves at a frequency close to the pendulum natural frequency in the absence of flow. The theoretical prediction of a natural frequency $\omega_0$ independent of the wave number is illustrated in red dashed line in fig.~\ref{fig:RD2D}a,b. For the small wind speed $U=1.16$~m/s, the prediction using constant frequency $\omega_0$ shows excellent agreement with the experimental values (fig.~\ref{fig:RD2D}a). At higher wind speed condition $U=3.16$~m/s, we observe a slight but significant increase of the resonant frequency with the wave number $kL$. This trend was also observed on the kinetic façades, and we attribute it to a wake-induced aerodynamic coupling between adjacent pendulums. In the following sections, we refer $\omega_0$ to be the overall resonant frequency that includes both the natural frequency and the weak coupling effects.

The branch II, on the other hand, corresponds to a continuum of Fourier modes located around a line $\omega = U_c k$, with a characteristic velocity $U_c$. This characteristic velocity increases with wind speed (fig.~\ref{fig:RD2D}), and can be interpreted as a convection speed of spatiotemporal structures along the pendulum chain. Compared to the field study, we have now access independently to the properties of the surrounding turbulent flow, as a function of the wind speed.

\begin{figure}[htbp]
  \includegraphics[width=0.9\columnwidth]{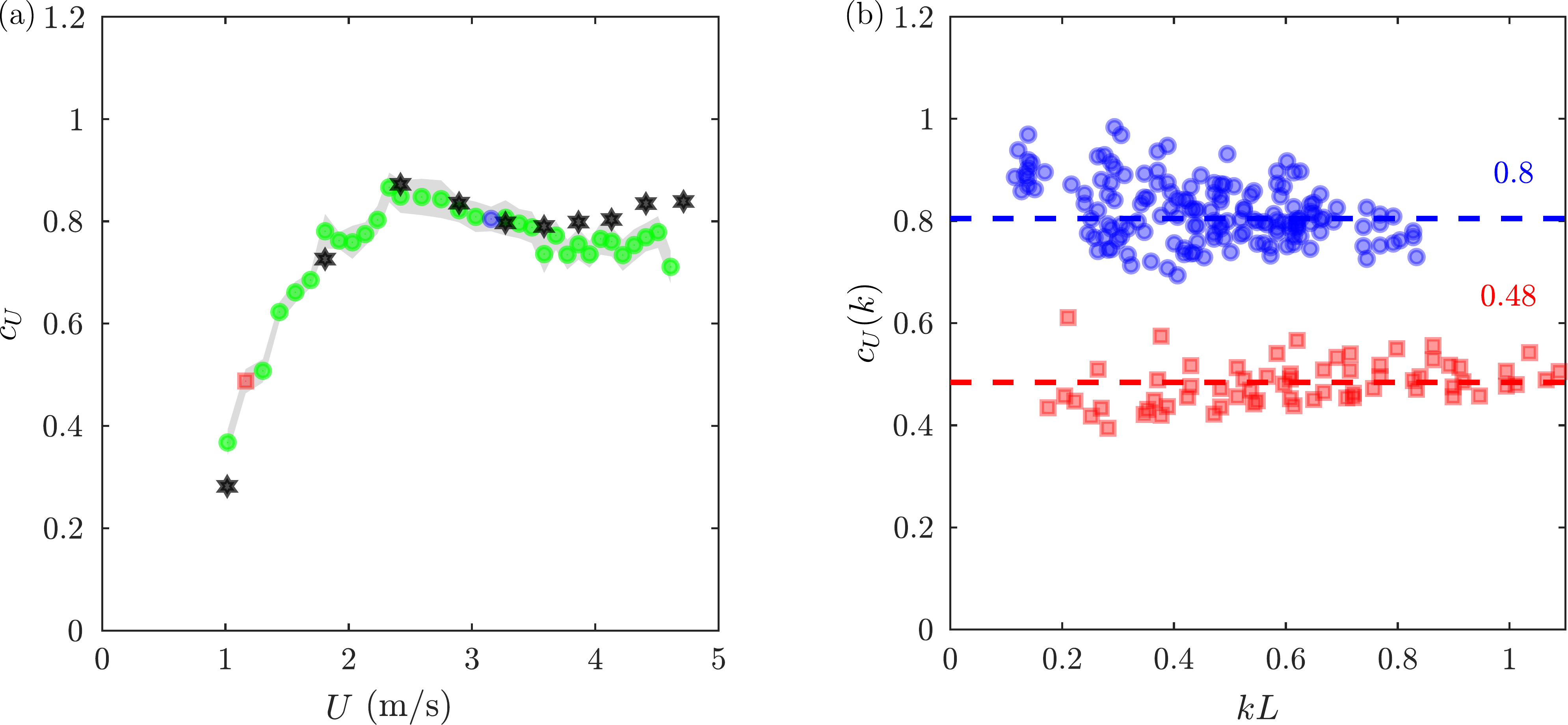}
  \caption{\label{fig:cv} a: mean normalized convection velocity as a function of mean wind velocity (green dots), data from 2-point correlation {with no pendulum} are shown for comparison (black hexagones). {Blue dots and red squares correspond to experiments of fig.~\ref{fig:RD2D} (red squares: $U=1.16$~m/s; blue dots: $U=3.16$~m/s).} The shaded area represents the rms of the linear fit error. b: local normalized convection velocity for two typical mean wind velocities {of fig.~\ref{fig:RD2D}} as a function of normalized wave number. }
\end{figure}

We then compare the convection speed of turbulent structures on the pendulum chain to the convection speed of pressure fluctuations in the absence of the pendulum chain. From the spatiotemporal spectrum of pendulum angles, for each wind speed, we apply a linear fit on the local maxima of $\hat{\theta}$ along the branch II. Doing so, we extract the slope of branch II as a function of the wind speed. To check that the convection speed $U_c$ scales with the wind speed $U$, we represent in fig.~\ref{fig:cv} the ratio {$c_U=U_c/U$} as a function of the wind speed $U$. This ratio increases with wind speed, and converge to a constant value {near} $0.8$ above $U>2$~m/s. This ratio is reminiscent of the convection speed of wall pressure fluctuations in a zero pressure gradient TBL~\cite{choi1990space,caiazzo2023effect}. 
To compare $U_c$ with the convection speed of pressure fluctuations in the turbulent channel, we performed measurements of two points spatiotemporal pressure fluctuations, in the absence of pendulum chain. The measurements were conducted with two acoustic pressure probes (PCB 103B01 with $\pm15\%$ sensibility) at the center line of the channel, spaced by $36$ cm which is a distance larger than the integral scale. We measured the convection speed of pressure fluctuations from temporal delay of the correlation peak between the two probe signals. The ratio {$c_U$} for pressure fluctuations have been superimposed on fig.~\ref{fig:cv} (black stars). We find a quantitative agreement with the convection speed of patterns along the pendulum chain, showing that branch II can indeed be interpreted as the signature of turbulent fluctuations traveling downstream, at a speed close but smaller than the wind speed. Note that the ratio {$c_U$}, however, can be sensitive to the type of turbulent flow, and we do not expect this curve to be universal. Another prediction from the literature of pressure fluctuations in TBL is the decrease of the convection speed with the wave number at large enough Reynolds numbers: the large modes (small $kL$) travel faster than the smaller ones, as observed numerically by Choi $\&$ Moin~\cite{choi1990space} and experimentally by Willmarth $\&$ Wooldridge~\cite{willmarth1962measurements}. From the motion of the pendulum chain, we have in theory access to the convection speed of patterns as a function of the wave number. In Fourier space, we extract the local slope of branch II as a function of the wave number $kL$. Figure~\ref{fig:cv}b shows the corresponding ratio {$c_U(k) = U_c(k)/U$} as a function of the wave number $kL$ for two wind speeds (blue and red). The scatter is due to the limited resolution in $k$. For small wind speed (red), we do not observe a significant trend with the wave number. For higher wind speed (blue), we see a slight decrease in the convection speed with the wave number, even though we are close to our limit of resolution. {Note that considering the pendulum's spacing $L$, the maximum dimensionless wave number $k_{max}L=2\pi$ introduces a wave number cutoff $k_{max} = 193~\mathrm{m^{-1}}$}:
 the structures smaller than $L$ will not be resolved by the chain oscillations.

The analysis of the two branches in Fourier space shows that the pendulum chain responds to the turbulent fluctuations as a collection of pendulums randomly pushed and pulled. To describe the pendulum angle statistics, we consider that the motion of each pendulum can still be described by an oscillator, and we model the acceleration exerted by the flow on the pendulum by an effective term $\Force_i(t)$:
\begin{equation}
  \partial_{tt}\theta_i + \Lambda \partial_t \theta_i + \Omega_0^2\theta_i= \Force_i(t),
  \label{eq:force_balance_full}
\end{equation}
where $\Lambda$ and $\Omega_0$ are respectively the damping coefficient and the resonant frequency of each pendulum in the presence of an external flow, that may depend on the wind speed. We have limited the analysis to the linearized equation, valid for small angles of oscillation. From the analysis of the spatiotemporal spectrum, we have seen that $\Omega_0=\omega_0$ holds for all wind speeds. For the sake of simplicity, we also assume no wind-induced coupling between adjacent pendulums. The acceleration $A_i$ is related to the normal stresses integrated over the plate surface by:
\begin{equation}
  \Force_i = \frac{1}{I_p}\int [{\bf \tau} {\bf n}] r ~ \textrm{d}S,
  \label{eq:effective_force}
\end{equation}
where $I_p = \rho_p w h l^3/3 = 10.9~$g cm$^2$ is the moment of inertia of the pendulum, ${\bf \tau}$ is the fluid stress tensor on the plate surface, ${\bf n}$ is the unit vector normal to the plate surface and the bracket $[\cdot]$ stands for the difference between the two plate surfaces. There is a priori no simple model for the forcing term $\Force_i$. Indeed, the presence of the pendulum chain introduces mixed boundary conditions at the plate surface, that will modify the flow structure near the plate. In the following, we will consider the plates as infinitely thin. Before investigating the pendulum chain case, it is interesting to briefly review two main limit cases: vanishing inertia (i) and rigid wall (ii). In limit (i), the plates do not modify the turbulent flow, and their motions follow the velocity fluctuations along the normal direction of the place. We then expect the normal stress difference $[{\bf \tau} {\bf n}]$ to scale with the instantaneous momentum flux perpendicular to the plate surface, {\textit{i.e.}}, $[{\bf \tau} {\bf n}] \sim \rho_a |u_\theta|u_\theta$, where $u_\theta$ is the velocity fluctuations along ${\bf n}$. From Taylor hypothesis~\cite{Pope_2000}, the velocity turbulent fluctuations travel at the mean speed of the flow, and the acceleration therefore scales as $\Force_i \sim 3 \rho_a/(2 \rho_p L h) u_\theta^2$. In limit (ii), the normal stress will be that of the wall-normal stress in growing turbulent boundary layers on both sides of the plate. The convection speed of the forcing would then be that of pressure fluctuations, which is typically smaller than the mean flow velocity, and the normal stress would scale as $\rho_a {u^*}^2$, where $u^*$ is the friction velocity. In practice, we can estimate if the pendulum freely follows the air, by looking at the magnitude of the relative pendulum velocity. At the tip of each plate, the pendulum velocity is of the order of $\theta_{rms} \omega_0 l$. 
\begin{figure}[htbp]
  \includegraphics[width=0.5\columnwidth]{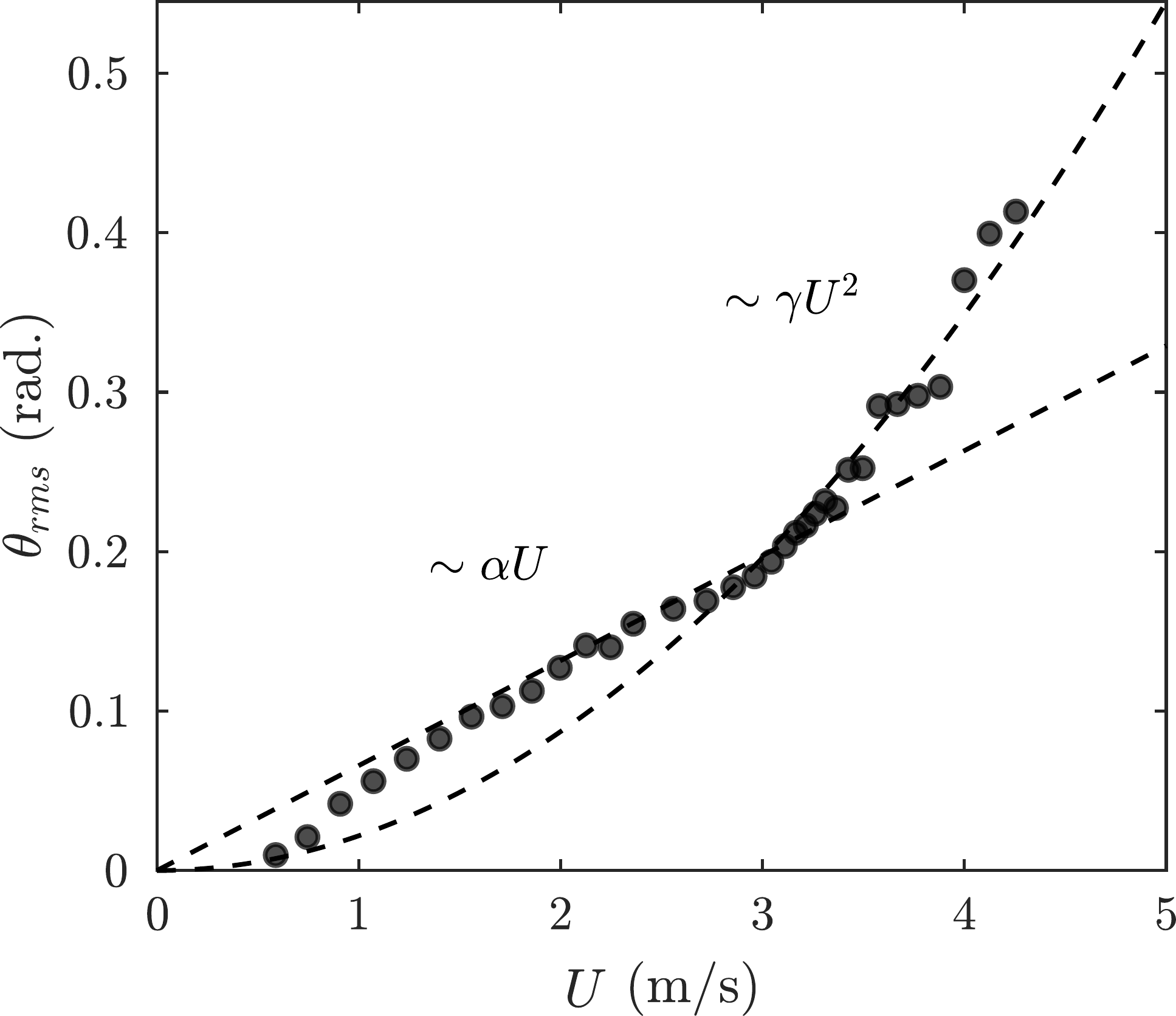}
  \caption{\label{fig:thetaRMS} root mean square $\theta_{rms}$ as a function of the wind speed. The dashed line represents the linear trend for $U<3$ m/s with a fitting coefficient $\alpha = 0.06~\mathrm{rad.m^{-1}s}$ and the dashed curve represents the quadratic variation for $U>3$ m/s with the coefficient $\gamma=2.18\times 10^{-2}~\mathrm{rad.m^{-2}s^2}$.}
\end{figure}

Figure~\ref{fig:thetaRMS} shows the root mean square $\theta_{rms}$ of the pendulum motion as a function of the wind speed. We observe that $\theta_{rms}$ is {roughly} proportional to the wind speed $U$ for $U < 3$~m/s, and increases faster than linear for $U \geq 3$~m/s. For $U<3$~m/s, a linear fit gives $\theta_{rms} = \alpha~U$ where $\alpha = 0.06~\pm~0.005~\mathrm{rad.m^{-1}s}$. To investigate if the pendulum surface moves with the flow velocity, we compare the azimuthal velocity at the plate tip $U_{plate}$  scales as $U_{plate} \sim \omega_0 l \theta_{rms} \sim 0.05 U$, which is of the order of but smaller than the velocity fluctuations $u' \sim 0.1 U$. The pendulum chain motion therefore lies in an intermediate regime, in which the plates can be approximated neither by a rigid wall nor a freely moving boundary. For $U>3$~m/s, we found a quadratic growth of $\theta_{rms}$  with a fitting coefficient $\gamma=2.18\times 10^{-2}~\mathrm{rad.m^{-2}s^2}$. 

 To understand the origin of these two scaling laws, we will perform a component-by-component analysis in Fourier space. 

\subsection{Resonant response in Fourier space} 

{A mechanism of spatio-temporal resonance between turbulent fluctuations and waves was introduced by Phillips~\cite{phillips1957generation} in the context of wind waves induced by turbulent wall-pressure fluctuations. More recently, Perrard \textit{et al.}~\cite{perrard2019turbulent,nove2020effect} combine component-by-component Fourier analysis with direct numerical simulation of wall pressure statistics to describe the statistics of surface waves below the onset of wind wave instability. A linear response theory was shown to describe accurately the observed behaviors. Here we proceed with a similar approach, to model the response of the pendulum chain in Fourier space.}
{Under the following assumptions:
\begin{itemize}
    \item A statistically stationary steady state is reached for both the turbulent flow and the pendulum chain oscillations.
    \item The angle of oscillations are small enough to describe the dynamics with a linear response theory.
    \item The pendulum oscillation dynamics is described by Eq.~\ref{eq:force_balance_full}.    
\end{itemize}
We Fourier transform in space and in time eq.~\ref{eq:force_balance_full} to express $\hat{\theta}(k,\omega)$ as:}
\begin{equation}
|\hat{\theta}|^2=\frac{|\hat{\Force}|^2}{(\omega^2-\omega_0^2(k))^2+\Lambda^2(k) \omega^2},
\label{eq:fourier}
\end{equation}
where $\hat{\mathcal{A}}$ is the Fourier transform of the effective acceleration, and the denominator is minimum at the pendulum resonant frequency $\omega = \omega_0$. The maximum response in Fourier space is therefore either located in the vicinity of the dispersion relation ($\omega \sim \omega_0$) where the denominator is minimum (branch I) or located where the forcing $\hat{\Force}$ is maximum (branch II). At a given wave number and an angular frequency $\omega \sim \omega_0$, the pendulum chain resonates with the spectral modes of $\hat{\mathcal{A}}$, and the amplitude growth is {only} limited by dissipative effects {which depends on the damping coefficient $\Lambda$}.

\begin{figure}[htbp]
  \includegraphics[width=1\columnwidth]{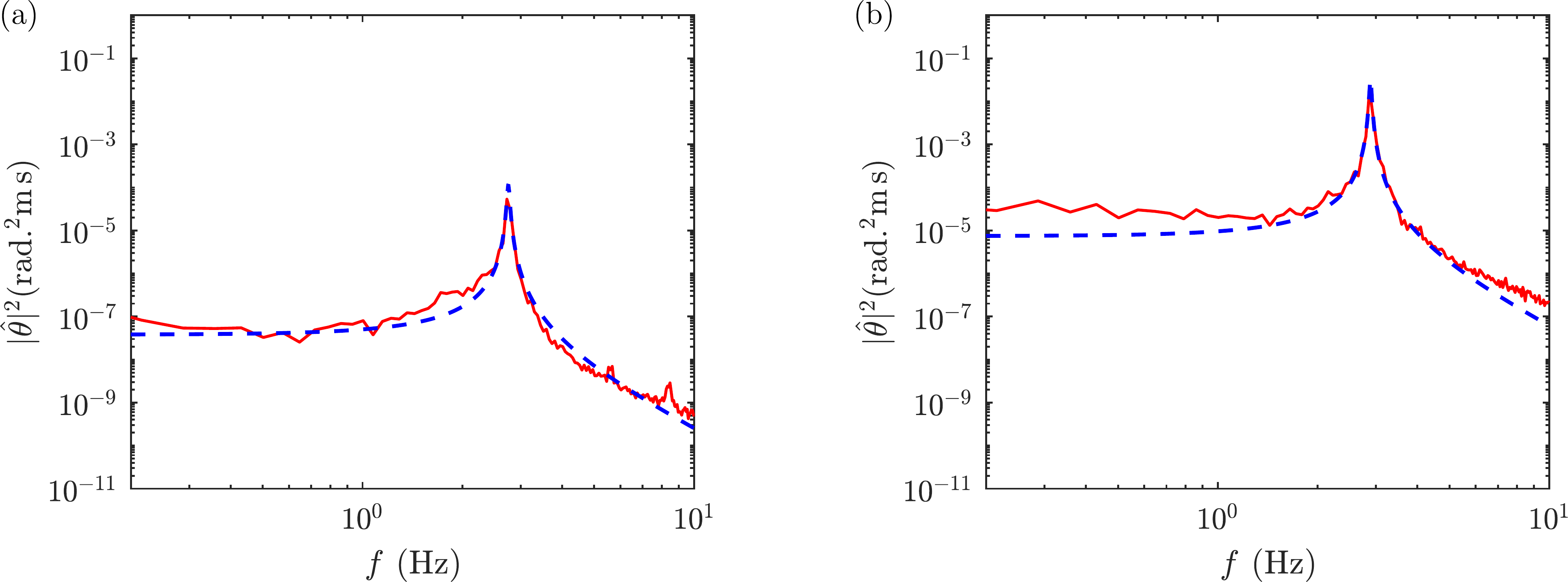}
  \caption{\label{fig:reson} Temporal spectrum $|\hat{\theta}|^2$ at $kL=0.75$ (red line) and the fit around the resonant frequency $f_r$ (blue dashed line). (a: $U=1.16$~m/s; b: $U=3.16$~m/s).}
\end{figure}
\begin{figure}
  \includegraphics[width=\columnwidth]{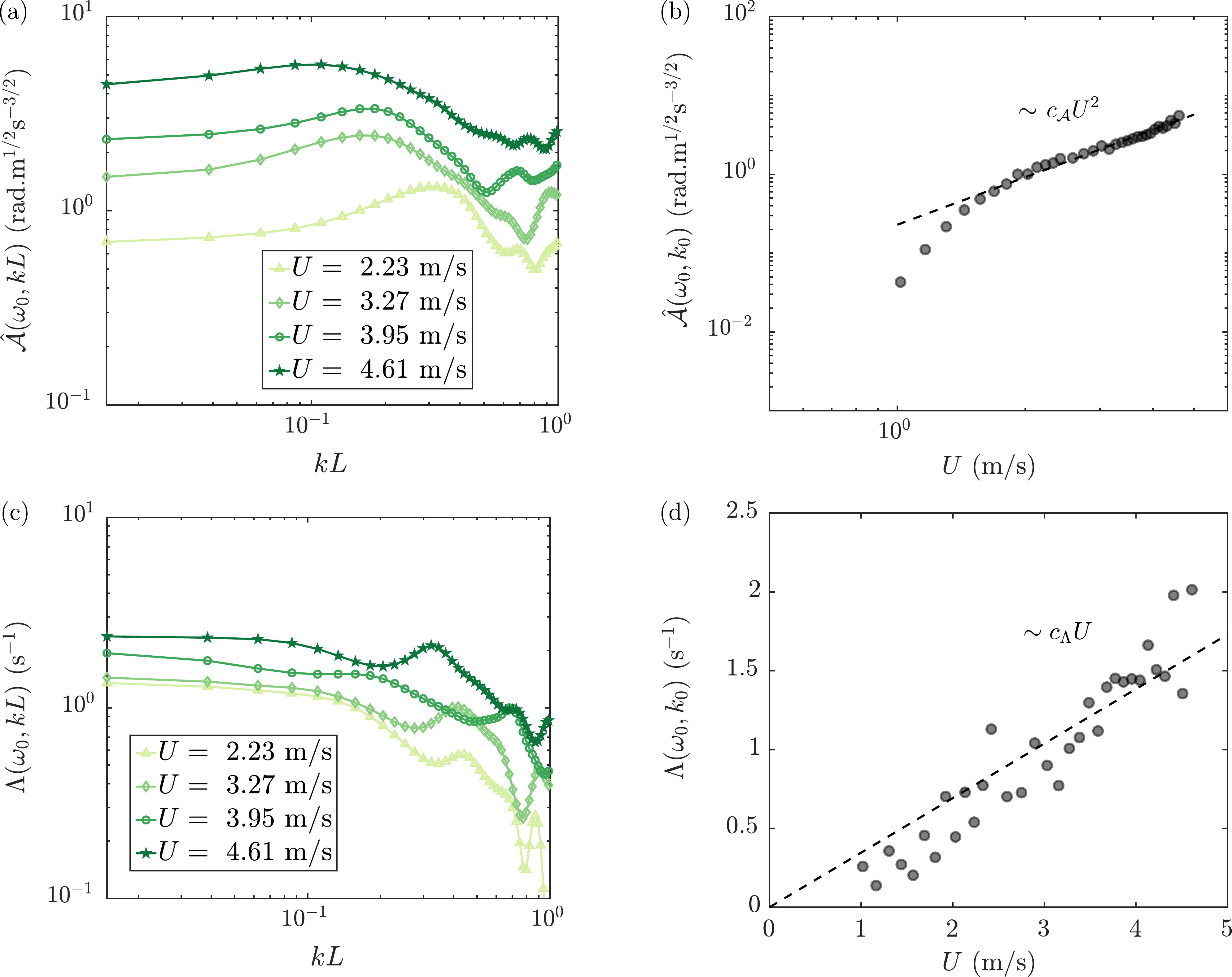}
  \caption{\label{fig:fitcoef} a: {fitted} forcing constant $\hat{\Force}(\omega_0, kL)$ as a function of the dimensionless wave number $kL$.  b: maximum forcing at the resonant peak $\hat{\Force}(\omega_0, k_0)$ as a quadratic function of the wind speed, with the fitting coefficient $c_\Force=0.23~\mathrm{rad.m^{-3/2}s^{1/2}}$. {c: fitted damping coefficient $\Lambda(\omega_0, kL)$ as a function of $kL$.}
  d: The fitted source term in Fourier space $\hat{\Force}(\omega_0, k_0)$ both at the resonant peak frequency $\omega_0$ and the maximum convection wave number $k_0=\omega_0/U_c$, as a function of the wind velocity. The black dashed line represents the linear scaling law $\propto U$ with a fitting coefficient $c_\Lambda=0.34~\mathrm{m^{-1}}$.}
\end{figure}

To test the validity of eq.~\ref{eq:fourier}, we compute {from the experimental data} the energy spectrum $|\hat{\theta}|^2$. Fig.~\ref{fig:reson}a,b shows the energy spectrum {for $kL=0.75$ as a function of the frequency $f$ for two different} wind speeds (red line). We observe resonant curves {near the resonant frequency $\omega_0$ in the absence of wind}, for all small wave numbers, $kL < 1$, and all wind speeds. To describe the shape of the resonant response near $\omega_0$, we fit the expression of eq.~\ref{eq:fourier} using two adjustable parameters, the effective {acceleration} $\hat{\Force}(\omega_0)$ at the resonance, and the damping coefficient $\Lambda$. We {then} neglect the variation of $\hat{\Force}$ with $\omega$, which is valid for sharp resonance, $\textit{i.e.}$ {for} $\Lambda \ll \omega_0$. We {eventually} find a quantitative agreement between eq.~\ref{eq:fourier} (blue dashed line) and the experimental measurements for $kL <1$ and all wind speeds. {From the fit of the resonant curve, we extract the values of $\hat{\Force}(\omega_0,kL)$ and $\Lambda$ as a function of the dimensionless wavenumber $kL$ and the wind speed.}
The effective acceleration $\hat{\Force}(\omega_0,kL)$ at the resonance is shown in fig.~\ref{fig:fitcoef}(a) as a function of the dimensionless wave number $kL$ for different wind speed (green color bar). The effective acceleration $\hat{\Force}(\omega_0,kL)$ exhibits a smooth maximum around $k_0 = \omega_0/U_c$, which corresponds to the cross-over between the branch I and II. The magnitude of $\hat{\Force}(\omega_0,kL)$ decreases both for smaller and larger wave numbers. Figure~\ref{fig:fitcoef}b shows the effective acceleration $\hat{\Force}_{max} = \hat{\Force}(\omega_0,L\omega_0/U_c)$, {located in Fourier space} at the intersection between branches I and II, as a function of the wind speed $U$. We find that the effective acceleration goes as $\hat{\Force}_{max} = c_\Force U^2$ with $c_\Force = 0.23~\mathrm{rad.m^{-3/2}s^{1/2}}$. {This effective acceleration originates from the torque of pressure fluctuations integrated over the plate surface (eq.~\ref{eq:effective_force}), which scales as $\hat{\Force}_{max} \sim U^2$ as expected for inertial forces}. 


{The fitted values of the damping coefficient $\Lambda$} are shown in fig.~\ref{fig:fitcoef}c as a function of the dimensionless wave number $kL$, for different wind speeds (color-coded). For small wave numbers ($kL<0.2$), the damping coefficient is almost constant and then decreases with $kL$. 
{Note that the fit is performed with a constant forcing term $\hat{\Force}$ which is a valid assumption only in the vicinity of the resonance.} {The damping coefficient  $\Lambda_{max} = \Lambda(\omega_0,L\omega_0/U_c)$ at the intersection between branches I and II is shown in figure~\ref{fig:fitcoef}d as a function of the wind speed $U$}. We find that $\Lambda_{max}$ increases linearly with the wind speed $U$. This increase of dissipation with the wind speed could either originate from an enhancement of the momentum transfer from the plate oscillation to the fluid by the turbulent flow, or by an increase of dissipation in the boundary layers, as observed recently for bubble oscillations in turbulence~\cite{riviere2024bubble}.


The maximum response is located in Fourier space at the intersection between the two branches. {From the values of the fitted coefficients $\Force$ and $\Lambda$}, we {now} estimate the contribution {of the resonant response near $\omega=\omega_0$ and $k_0 = \omega_0/U_c$ to $\theta_{rms}$}. To do so, we consider the integral of eq.~\ref{eq:fourier} over the wave numbers $k$ and the angular frequency $\omega$:
\begin{equation}
\theta_{rms}^2 = \int \dd k \int \dd \omega  \frac{|\hat{\Force}|^2}{(\omega^2-\omega_0^2(k))^2+\Lambda^2(k) \omega^2}.
\label{eq:theta_integral}
\end{equation}


We estimate the resonant response $\theta_r$ by considering wave numbers in the range $k \in [\omega_0/U_c-\Delta k/2,~\omega_0/U_c+\Delta k/2]$, {where the spectral width $\Delta k \sim \omega_0/U_c$ is associated to the typical fluctuations of the} convection speed of turbulent structures. In the limit of large quality factor $\omega_0/\Lambda \gg 1$, we approximate the numerator by its value at {the} resonance, $\hat{\Force} = \hat{\Force}(\omega_0, \omega_0/U_c)$, and we obtain an estimate of the resonant response as:
\begin{equation}
\theta_{r}^2 = \frac{2 \Delta k {c_U}^2 U^4}{\omega_0^3} \int_{-\infty}^{+\infty} \dd \tomega  \frac{1}{(\tomega^2-1)^2+\Lambda^2(k)/\tomega_0^2~\tomega^2},
\label{eq:theta_resonance}
\end{equation}
where the integral $I(\Lambda) = \int \dd \tomega~ ((\tomega^2-1)^2+\Lambda^2 \tomega^2)^{-1} = \pi/\Lambda$ with $\tomega$ the normalised frequency. We eventually obtain an estimate for the resonant response :
\begin{equation}
\theta_{r}^2 = \frac{2 c_\Force^2 U^4}{\omega_0^2 U_c}\pi \frac{\omega_0}{\Lambda}.
\label{eq:thetar_scaling}
\end{equation}
{Using the fitted expression $\Lambda = c_\Lambda U$ and the expression of the advection speed $U_c = c_U U$, we eventually obtain an expression for the resonant response}:
\begin{equation}
\theta_{r} = \sqrt{\frac{2 \pi c_\Force^2}{c_\Lambda c_U \omega_0}} U.
\label{eq:thetar_scaling}
w\end{equation}
We eventually found that the resonant response {yields an amplitude of oscillation proportional to the wind speed $U$}, as observed experimentally for $U<3$~m/s. Using the fitted values  $c_{\Lambda}$,$c_\Force$ and $c_U\sim=0.8$, we obtain $\theta_r = c_\theta U$ with $c_\theta = 0.23~\mathrm{rad.m^{-1}s}$. This coefficient is of the same order of magnitude but larger than the fitted value of the slope $\alpha = 0.06~\mathrm{rad.m^{-1}s}$ shown in fig.~\ref{fig:thetaRMS}. 
Note that at higher wind speeds, {the damping coefficient increases and the assumption of sharp resonance is not fulfilled. Consequently, the pendulum response far from the resonance cannot be neglected. The integral over Fourier space of Eq.~\ref{eq:theta_integral} is eventually dominated by the energy along branch II}. Assuming a constant forcing term {along the line $\omega = k U_c$} for $k L_{int} <1$, the amplitude of pendulum oscillation $\theta_{rms}$ given by eq.~\ref{eq:theta_integral} {scales} quadratically with $U$, as $\theta_{rms} \sim c_\Force U^2/(\omega_0^{3/2}L_{int}^{1/2})$. 
Using the fitted coefficient $c_\Force$ and a constant integral length $L_{int} = 5$ cm for large wind speeds, we obtain $\gamma_{th} = 1.41\times 10^{-2}~\mathrm{rad.m^{-2}s^2}$, which is in fair agreement with $\gamma_{exp}$ previously found for the $\theta_{rms}$ fit in the quadratic regime. 
To summarize, we find that in the explored range of wind speed, we observed two regimes of pendulum response. At low wind speed, the oscillations are dominated by the resonant response to pressure fluctuations, while at higher wind speed, the response is dominated by the direct forcing from the pressure fluctuations traveling along the pendulum chain.

\section{Conclusion} \label{sec:conclusion} Inspired by Ned Kahn's kinetic façade artwork, we conducted a qualitative analysis of structures propagating on the building facades. To explain the physical processes at play, we studied a one dimensional chain of weakly coupled pendulums immersed in a turbulent flow. We performed analysis in Fourier space, and we showed that both the natural system and the laboratory analog exhibit energy along two main branches in Fourier space. These two branches correspond to two different mechanisms. The branch I corresponds to the resonant response of each pendulum at its natural oscillation frequency. We measured the associated damping rate as a function of the wind speed, as well as the magnitude of the forcing at the resonance. We showed that the dissipation is proportional to the wind speed. The forcing term increases as $U^2$ with the wind speed, as expected from inertial forces. For small damping ($U<3$~m/s for the laboratory analog), the response is dominated by this mechanism, and the oscillation amplitude $\theta_{rms}$ scales as the wind speed $U$. The second mechanism (branch II) corresponds to the response to turbulent fluctuations traveling downstream along the wire at an almost constant convection speed independent of the wave number. This convection speed turns out to be equal to the convection speed of pressure fluctuations measured with pressure probes in the absence of the pendulum chain. This convection speed is of the order of the wind speed, but smaller (typically 80\% of wind speed). This pendulum system, either in one dimension or two dimensions, naturally responds at large wave number and small frequencies, revealing some large scale structures of the flow. However, the resonant response generates a filter, which preferentially amplifies the fluctuations at the pendulum frequency.

\begin{acknowledgments}

We acknowledge fruitful discussions with Ned Kahn, Antonin Eddi, Laurette Tuckerman, Guowei He, Marc Rabaud, Frédéric Moisy, Aliénor Rivière, and Philippe Bourianne. We are specially grateful to Amaury Fourgeaud for technical support, Gauthier Bertrand for the wind tunnel design, and Marc Fermigier for scientific support. We also acknowledge Jörg Moor from \textit{Swiss Science Center Technorama} and John Gray from \textit{Dundee City Council} for providing Ned Kahn’s facade dimensions. This work was supported by a PSL Junior Fellow Starting Grant 2022 (No. 2022–305) and by the Agence Nationale de la Recherche with grants ANR Lascaturb (reference ANR-23-CE30-0043).

\end{acknowledgments}

\bibliographystyle{unsrt} 
\bibliography{biblioWV.bib} 

\end{document}